\documentclass{revtex4-2}%
\usepackage[latin9]{inputenc}
\usepackage{color}
\usepackage{amsmath}
\usepackage{amssymb}
\usepackage{graphicx}
\usepackage[unicode=true,  bookmarks=false,  breaklinks=false,pdfborder={0 0
1},backref=section,colorlinks=false]{hyperref}
\usepackage{amsfonts}
\usepackage{subfigure}
\usepackage{cleveref}
\usepackage{listings}
\usepackage{xcolor}
\usepackage{array}
\usepackage{url}%
\hypersetup{
colorlinks,linkcolor=blue,citecolor=blue,urlcolor=blue}
\makeatletter
\@ifundefined{textcolor}{}{
\definecolor{BLACK}{gray}{0}
\definecolor{WHITE}{gray}{1}
\definecolor{RED}{rgb}{1,0,0}
\definecolor{GREEN}{rgb}{0,1,0}
\definecolor{BLUE}{rgb}{0,0,1}
\definecolor{CYAN}{cmyk}{1,0,0,0}
\definecolor{MAGENTA}{cmyk}{0,1,0,0}
\definecolor{YELLOW}{cmyk}{0,0,1,0}
}

\makeatother
\begin{document}
\preprint{CTP-SCU/2021032}
\title{Holographic Superconductors in a Non-minimally Coupled Einstein-Maxwell-scalar Model}
\author{Yiqian Chen$^{a}$}
\email{chenyiqian@stu.scu.edu.cn}
\author{Xiaobo Guo$^{b}$}
\email{guoxiaobo@czu.edu.cn}
\author{Peng Wang$^{a}$}
\email{pengw@scu.edu.cn}
\affiliation{$^{a}$Center for Theoretical Physics, College of Physics, Sichuan University,
Chengdu, 610064, China}
\affiliation{$^{b}$Mechanical and Electrical Engineering School, Chizhou University,
Chizhou, Anhui, 247000, PR China}

\begin{abstract}
In this paper, we investigate holographic superconductors dual to
asymptotically anti-de Sitter black holes in an Einstein-Maxwell-scalar model
with a non-minimal coupling between the scalar and Maxwell fields. In the
probe limit, it shows that the scalar condensate occurs below the critical
temperature $T_{c}$, and decreases with the increase of the coupling constant
$\alpha$. On the other hand, the the critical temperature $T_{c}$ increases as
the coupling constant $\alpha$ grows. We also calculate the optical
conductivity of the holographic superconductor, and observe that a gap forms
below $T_{c}$. Interestingly, the non-minimal coupling can lead to a spike
occurring in the gap at a low temperature.

\end{abstract}
\maketitle
\tableofcontents

\section{Introduction}

\label{sec:Introduction}

Since superconductivity was first discovered in 1911 by Heike Kamerlingh Onnes
\cite{Delft2010}, people have paid great attention to the research of
superconducting theory and superconducting materials. In 1950, Landau and
Ginzburg proposed a theory to describe the superconductivity by a second order
phase transition, which is the famous Ginzburg--Landau theory
\cite{Ginzburg:1950sr}. Seven years later, a more complete microscopic theory
of superconductivity was proposed by Bardeen, Cooper and Schrieffer, which is
known as BCS theory \cite{Bardeen1957,Bardeen1957a}. Meanwhile,
superconducting materials have also been developing rapidly. For example, a
room-temperature superconductor was made at 267 GPa in 2020 \cite{Snider2020}.
However, the present theories are not complete enough to describe the high
temperature superconductivity since our understanding of strong correlation
physics is still superficial.

A remarkable conjecture of string theory was discovered at the end of last
century, i.e., the AdS/CFT correspondence, which states that a string theory
on $AdS_{5}$ is equivalent to a $\mathcal{N}=4$ super-Yang-Mills theory in
4-dimensional spacetime \cite{Maldacena:1997re}. The conjecture was soon
extended to the gauge/gravity correspondence and holographic principle. One of
the most powerful feature of the AdS/CFT correspondence is that it describes a
strong-weak duality, which provides a tool to understand a strongly
interacting gauge theory by studying a dual weakly interacting gravitational
theory. Therefore, it is natural to study the superconductivity by the AdS/CFT
correspondence. Inspired by the observation that the spontaneous $U(1)$
symmetry breaking of the order parameter in the Ginzburg--Landau theory leads
to the superconductivity, interestingly, a similar mechanism acting on the
scalar field in the bulk was proposed in \cite{Gubser:2008px}, which results
in the scalarization of black holes. In particular, a theoretical model of
holographic superconductors was built by the authors of \cite{Hartnoll:2008vx}%
, which describes a (2+1)-dimensional s-wave superconductor. Specifically,
they considered the Abelian-Higgs model in the probe limit and found that the
superconducting phase transition is dual to the second order phase transition
in the bulk. To be more realistic, the backreactions of the scalar and
electromagnetic fields were considered \cite{Hartnoll:2008kx}, and the
(2+1)-dimensional superconductor model was generalized to (3+1)-dimensional
spacetime \cite{Horowitz:2008bn,Brihaye:2010mr}. One can also construct p-wave
and d-wave holographic superconductors by introducing more matter fields
\cite{Gubser2008,Cai:2010cv,Chen:2010mk,Benini:2010pr,Kim:2013oba,Cai:2013aca}%
. Moreover, many interesting works have been done in the past decade, for
instance, the holographic superconductor models in present of the dynamical gauge field
\cite{Domenech:2010nf,Montull:2012fy,Salvio2012}, the nonlinear electrodynamics
\cite{Jing:2010zp,Jing:2011vz,Gangopadhyay:2012am,Gangopadhyay:2012np,Mu:2017usw,Wang:2018hwg}%
, the Gauss-Bonnet corrections
\cite{Gregory:2009fj,Pan:2009xa,Pan:2010at,Cai:2010cv}, the external magnetic
field \cite{Albash:2008eh,Ge:2010aa}. Besides, the related contents have been
extensively investigated, e.g., the flavor
\cite{Ammon:2008fc,Ammon:2009fe,Kaminski:2010zu}, the vortex
\cite{Albash:2009iq,Dias:2013bwa}, the hydrodynamics \cite{Amado:2009ts}, the
entanglement entropy \cite{Albash:2012pd}, the zero temperature limit
\cite{Horowitz:2009ij,Nishioka:2009zj,Konoplya:2009hv}, the Lifshitz scaling
\cite{Brynjolfsson:2009ct,Lu:2013tza} and the Josephson Junctions
\cite{Horowitz:2011dz,Wang:2012yj,Hu:2015dnl}.

On the other hand, the phenomenon of spontaneous scalarization has attracted
great attention since it was first discovered for neutron stars in
scalar-tensor models. Later, this phenomenon was extended to black holes
\cite{Damour:1993hw,Cardoso:2013opa,Cardoso:2013fwa}. Recently, scalarized
black holes solutions have been found in the extended
Scalar-Tensor-Gauss-Bonnet (eSTGB) gravity, in which the scalar field
non-minimally couples to the Ricci scalar and the Gauss-Bonnet term
\cite{Doneva:2017bvd,Silva:2017uqg,Antoniou:2017acq,Cunha:2019dwb,Herdeiro:2020wei,Berti:2020kgk}%
. The non-minimal coupling can provide an effective mass for the scalar field
and lead to the spontaneous scalarization, which can be interpreted as the
holographic phase transition \cite{Brihaye:2019dck}. To have a better
understanding of the evolution of spontaneous scalarization, it is convenient
to study simpler Einstein-Maxwell-scalar (EMS) models with a non-minimal
coupling between the scalar and Maxwell fields \cite{Herdeiro:2018wub}.
Various non-minimal coupling functions and properties of the EMS models have
been studied in
\cite{Fernandes:2019rez,Blazquez-Salcedo:2020nhs,Astefanesei:2019pfq,Fernandes:2019kmh,Zou:2019bpt,Fernandes:2020gay,Peng:2019cmm,Myung:2018vug,Myung:2019oua,Zou:2020zxq,Myung:2020etf,Mai:2020sac,Astefanesei:2020qxk,Myung:2018jvi,Blazquez-Salcedo:2020jee,Myung:2020dqt,Myung:2020ctt,Guo:2020zqm,Wang:2020ohb,Konoplya:2019goy,Brihaye:2019gla,Hod:2020ljo,Hod:2020ius,Hod:2020cal,Mahapatra:2020wym,Wang:2020ohb,Gan:2021xdl,Gan:2021pwu}
. Therefore, it is of great interest to investigate the holography for
asymptotically anti-de Sitter black holes in the EMS model.

The rest of the paper is organized as follows. In section
\ref{sec:Holography in EMS Model}, we briefly introduce the EMS models with a
non-minimal coupling between the scalar and Maxwell fields in the
asymptotically AdS spacetime, and discuss the correspondence between
asymptotic forms of bulk fields and physical quantities of dual CFT in the
probe limit. In section \ref{sec:Numeric-Results}, we present and discuss the
numerical results of the condensates and the optical conductivity. Finally, we
conclude with a brief discuss in section \ref{sec:Conclusion}.

\section{Holography in EMS Model}

\label{sec:Holography in EMS Model}

In this section, we study the EMS model coupled to a charged scalar field in
the asymptotically AdS spacetime. The action of the EMS model in the
4-dimensional spacetime is
\begin{equation}
S_{\text{bulk}}=\frac{1}{16\pi G_{N}}\int d^{4}x\sqrt{-g}\left[  R+\frac
{6}{L^{2}}-\left\vert \nabla\Psi^{2}-iqA\Psi\right\vert ^{2}-m^{2}\left\vert
\Psi\right\vert ^{2}-\frac{h\left(  \Psi\right)  }{4}F_{\mu\nu}F^{\mu\nu
}\right]  , \label{eq:action}%
\end{equation}
where we take the Newton's constant $G_{N}=1$ for simplicity throughout this
paper. In the action $\left(  \ref{eq:action}\right)  $, the complex scalar
field $\Psi$ has mass $m$ and non-minimally coupled to the gauge field
$A_{\mu}$ with charge $q$, $F_{\mu\nu}=\partial_{\mu}A_{\nu}-\partial_{\nu
}A_{\mu}$ is the electromagnetic field strength tensor, $h\left(  \Psi\right)
$ is the non-minimal coupling function of the scalar and the gauge fields, and
$L$ is the curvature radius of AdS spacetime. In this paper, we focus on the
coupling function $h\left(  \Psi\right)  =e^{\alpha\Psi^{2}}$ with $\alpha
\geq0$.

If one rescales $A_{\mu}=\widetilde{A}_{\mu}/q$, $\Psi=\widetilde{\Psi}/q$ and
$\alpha=\widetilde{\alpha}q^{2}$, the matter part of the action $\left(
\ref{eq:action}\right)  $ has a factor $1/q^{2}$. Holding $\widetilde{A}_{\mu
}$, $\widetilde{\Psi}$ and $\widetilde{\alpha}$ fixed with $q\rightarrow
\infty$ gives the probe limit. In the probe limit, the rescaled scalar and
electromagnetic fields do not backreact the background while the interactions
between the scalar and electromagnetic fields are still retained. For
simplicity, we do not mark tilde on rescaled quantities in what fallows. We
now consider the probe scalar field $\Psi$ and the electromagnetic field
$A_{\mu}$ in the background of a planar Schwarzschild-AdS black hole solution
with the metric,%
\begin{equation}
ds^{2}=-f\left(  r\right)  dt^{2}+\frac{1}{f\left(  r\right)  }dr^{2}%
+r^{2}\left(  dx^{2}+dy^{2}\right)  ,
\end{equation}
where the metric function $f\left(  r\right)  $ is
\begin{equation}
f\left(  r\right)  =\frac{r^{2}}{L^{2}}-\frac{M}{r},
\end{equation}
and $M$ is the black hole mass. The Schwarzschild-AdS black hole has the event
horizon at $r_{0}=M^{\frac{1}{3}}L^{\frac{2}{3}}$, and its temperature is
$T=\frac{3M^{1/3}}{4\pi L^{4/3}}$.

\subsection{Condensates of the scalar field}

Varying the action $\left(  \ref{eq:action}\right)  $ with respect to the
scalar field $\Psi$ and the gauge field $A_{\mu}$, one obtains the equations
of motion,
\begin{align}
\nabla_{\mu}\nabla^{\mu}\Psi-\left(  A_{\mu}A^{\mu}+m^{2}\right)  \Psi
-\frac{\alpha}{4}e^{\alpha\Psi^{2}}F^{2} &  =0,\nonumber\\
2\Psi^{2}A_{\mu}-\nabla^{\nu}\left(  e^{\alpha\Psi^{2}}F_{\nu\mu}\right)   &
=0.\label{eq:EOMs}%
\end{align}
In the following, we consider a planer symmetric ansatz for the scalar field
and the gauge field,%
\begin{equation}
\Psi=\Psi\left(  r\right)  \text{ and }A_{\mu}dx^{\mu}=\phi\left(  r\right)
dt.\label{eq:ansatz}%
\end{equation}
Plugging the above ansatz into the equations of motion $\left(  \ref{eq:EOMs}%
\right)  $ yields
\begin{align}
\Psi^{\prime\prime}+\left(  \frac{f^{\prime}}{f}+\frac{2}{r}\right)
\Psi^{\prime}+\frac{q^{2}\phi^{2}}{f^{2}}\Psi-\frac{m^{2}}{f}\Psi+\frac
{\alpha\phi^{\prime2}}{2f}e^{\alpha\Psi^{2}}\Psi &  =0,\nonumber\\
\phi^{\prime\prime}+\left(  2\alpha\Psi\Psi^{\prime}+\frac{2}{r}\right)
\phi^{\prime}-\frac{2q^{2}\Psi^{2}e^{-\alpha\Psi^{2}}}{f}\phi &
=0,\label{eq:ansatz's EOM}%
\end{align}
where primes denote the derivatives with respect to the radial coordinate $r$.

Solving the equations of motion $\left(  \ref{eq:ansatz's EOM}\right)  $ at
the infinite boundary $r\rightarrow\infty$, these solutions behave as
\begin{align}
\Psi &  =\frac{\Psi^{-}}{r^{\Delta_{-}}}+\frac{\Psi^{+}}{r^{\Delta_{+}}%
}+\cdots,\nonumber\\
\phi &  =\mu-\frac{\rho}{r}+\cdots, \label{eq:asymptotic form}%
\end{align}
where $\Delta_{\pm}=\frac{3\pm\sqrt{9+4m^{2}L^{2}}}{2}$, $\mu$ is the chemical
potential and $\rho$ is the charge density in the boundary theory. In this
paper, we take $m^{2}=-\frac{2}{L^{2}}$, which is commonly used in the
$AdS_{4}/CFT_{3}$ correspondence. Therefore, the asymptotic expansions
$\left(  \ref{eq:asymptotic form}\right)  $ reduce to
\begin{equation}
\Psi=\frac{\Psi^{(1)}}{r^{1}}+\frac{\Psi^{(2)}}{r^{2}}+\cdots.
\label{eq:asymptotic form2}%
\end{equation}
According to the $AdS_{4}/CFT_{3}$ correspondence, one can choose a falloff
solution with $\Psi^{(1)}=0$ or $\Psi^{(2)}=0$, which means the condensate
turns on without being sourced. The remainder nonzero $\Psi^{(i)}$ can be read
as the expectation value of the dual operator $\mathcal{O}_{1}$ or
$\mathcal{O}_{2}$ in the dual CFT, which is given by
\begin{equation}
\left\langle \mathcal{O}_{i}\right\rangle =\sqrt{2}\Psi^{(i)},\epsilon
_{ij}\Psi^{(i)}=0,i=1,2. \label{eq:expectation value}%
\end{equation}
Here, the factor $\sqrt{2}$ is a convenient normalization. In short, the
condensate $\left\langle \mathcal{O}_{i}\right\rangle $, the chemical
potential $\mu$ and the charge density $\rho$ can be determined by solving the
equations of motion $\left(  \ref{eq:ansatz's EOM}\right)  $ with a proper
boundary condition.

\subsection{Conductivity}

Based on solutions of the equations $\left(  \ref{eq:ansatz's EOM}\right)  $,
one can compute the conductivity in the dual CFT by solving the fluctuations
of the vector potential $A_{x}$ in the bulk. The fluctuation $A_{x}$ with a
time dependent form $e^{-i\omega t}$ obeys the zero spatial momentum
electromagnetic equation%
\begin{equation}
A_{x}^{\prime\prime}+\left(  2\alpha\psi\psi^{\prime}+\frac{f^{\prime}}%
{f}\right)  A_{x}^{\prime}+\left(  \frac{\omega^{2}}{f^{2}}-\frac{2\psi
^{2}e^{-\alpha\psi^{2}}}{f}\right)  A_{x}=0. \label{eq:fluctuations' eq}%
\end{equation}
To solve this perturbed equation, we impose the ingoing wave boundary
condition at the horizon for causal propagation on the boundary, i.e.,
$A_{x}\propto f^{-i\omega/3r_{0}}|_{r\rightarrow r_{0}}$. On the other hand,
the asymptotic behavior of the fluctuation at a large radius is given by%
\begin{equation}
A_{x}=A_{x}^{(0)}+\frac{A_{x}^{(1)}}{r}+\cdots.
\label{eq:asymptotic form of fluctuations}%
\end{equation}
According to the AdS/CFT dictionary, the dual source and expectation value for
the electric field are given by%
\begin{equation}
E_{x}=-\dot{A}_{x}^{(0)}=i\omega A_{x}^{(0)},\left\langle J_{x}\right\rangle
=A_{x}^{(1)}, \label{eq:source and expectation value}%
\end{equation}
respectively. Then we can obtain the conductivity by Ohm's law%
\begin{equation}
\sigma\left(  \omega\right)  =-\frac{iA_{x}^{(1)}}{\omega A_{x}^{(0)}}.
\label{eq:conductivity}%
\end{equation}

\section{Numerical Results}

\label{sec:Numeric-Results}

In this section, we present the numerical results, e.g., the condensate as a
function of temperature and the properties of the optical conductivity. When
performing numerical calculations, one can set $L=1$ and $r_{0}=1$ by using
two scaling symmetries of the equations of motion
\cite{Gubser:2008px,Hartnoll:2008kx},
\begin{equation}%
\begin{array}
[c]{ccc}%
r\rightarrow ar, & t\rightarrow at, & L\rightarrow aL,
\end{array}
\end{equation}
and
\begin{equation}%
\begin{array}
[c]{ccc}%
r\rightarrow ar, & \left(  t,x,y\right)  \rightarrow\left(  t,x,y\right)
/a, & \phi\rightarrow a\phi.
\end{array}
\label{eq:scaling sym2}%
\end{equation}
Note that the temperature $T$ has mass dimension one, the chemical potential
$\mu$ has mass dimension one, the charge density $\rho$ has mass dimension
two, and the condensates $\left\langle \mathcal{O}_{1}\right\rangle $ and
$\left\langle \mathcal{O}_{2}\right\rangle $ have mass dimension one and two,
respectively. Usually, one can rescale quantities of interest with the
chemical potential $\mu$ in a grand canonical ensemble or the charge density
$\rho$ in a canonical ensemble. In this paper, we consider a canonical
ensemble and hence introduce the following rescaled quantities%
\begin{equation}
\widetilde{\left\langle \mathcal{O}_{1}\right\rangle }=\frac{\left\langle
\mathcal{O}_{1}\right\rangle }{\sqrt{\rho}},\widetilde{\left\langle
\mathcal{O}_{2}\right\rangle }=\frac{\sqrt{\left\langle \mathcal{O}%
_{2}\right\rangle }}{\sqrt{\rho}},\tilde{T}=\frac{T}{\sqrt{\rho}},\tilde
{T}_{c}=\frac{T_{c}}{\sqrt{\rho_{c}}}.
\end{equation}
For simplicity we will omit the tilde notation for the above quantities in the
remainder of the section.

\subsection{Condensate}

\begin{figure}[ptb]
\includegraphics{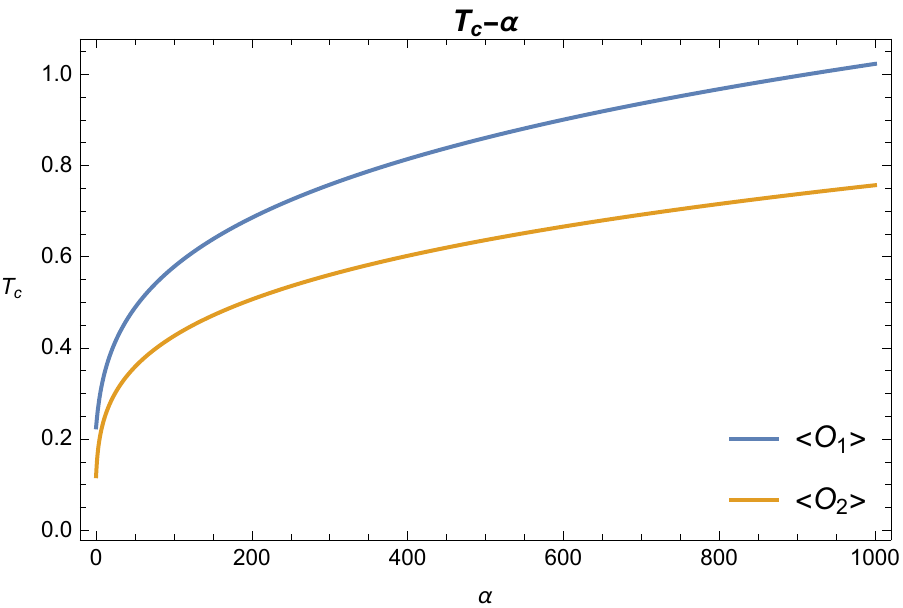}\caption{The critical temperature $T_{c}$ as a
function of the coupling $\alpha$ for the condensates $\left\langle
\mathcal{O}_{1}\right\rangle $ and $\left\langle \mathcal{O}_{2}\right\rangle
$.}%
\label{Fig1}%
\end{figure}

In this subsection, we numerically solve the equations $(\ref{eq:ansatz's EOM}%
)$ with the boundary conditions discussed in Section
\ref{sec:Holography in EMS Model}. We find that there exists a critical
temperature $T_{c}$ associated with the holographic superconducting phase
transition in the dual CFT. Above the critical temperature the condensate is
zero while below the temperature the condensate occurs and the superconducting
state in the boundary forms. We plot the critical temperature $T_{c}$ as a
function of the coupling $\alpha$ in Fig. \ref{Fig1} and show that $T_{c}$
increases as $\alpha$ increases, which means that the non-minimal coupling of
the scalar and the electromagnetic fields make the occurrence of the
condensate easier. In other words, holographic superconductors have a higher
superconducting transition temperature in the model with a larger coupling
$\alpha$.

\begin{figure}[ptb]
\includegraphics[width=8cm]{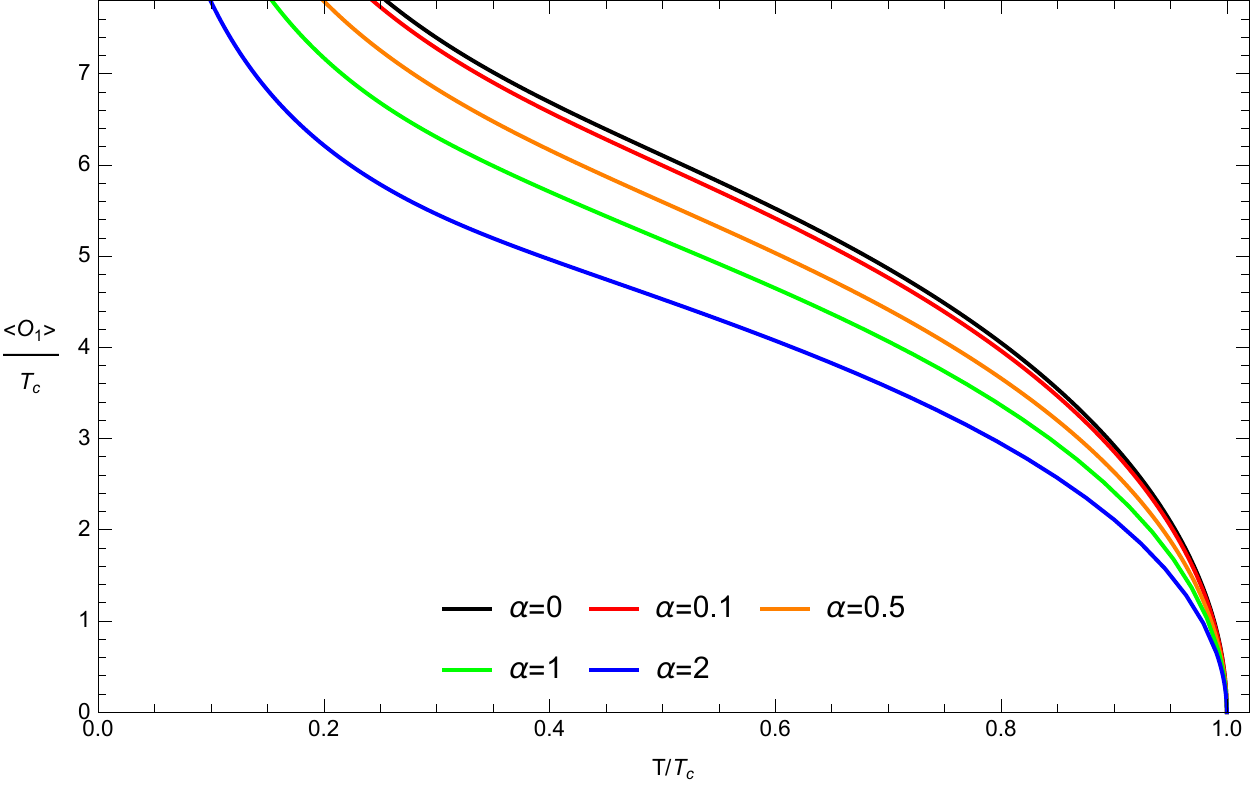}~~~~~\includegraphics[width=8cm]{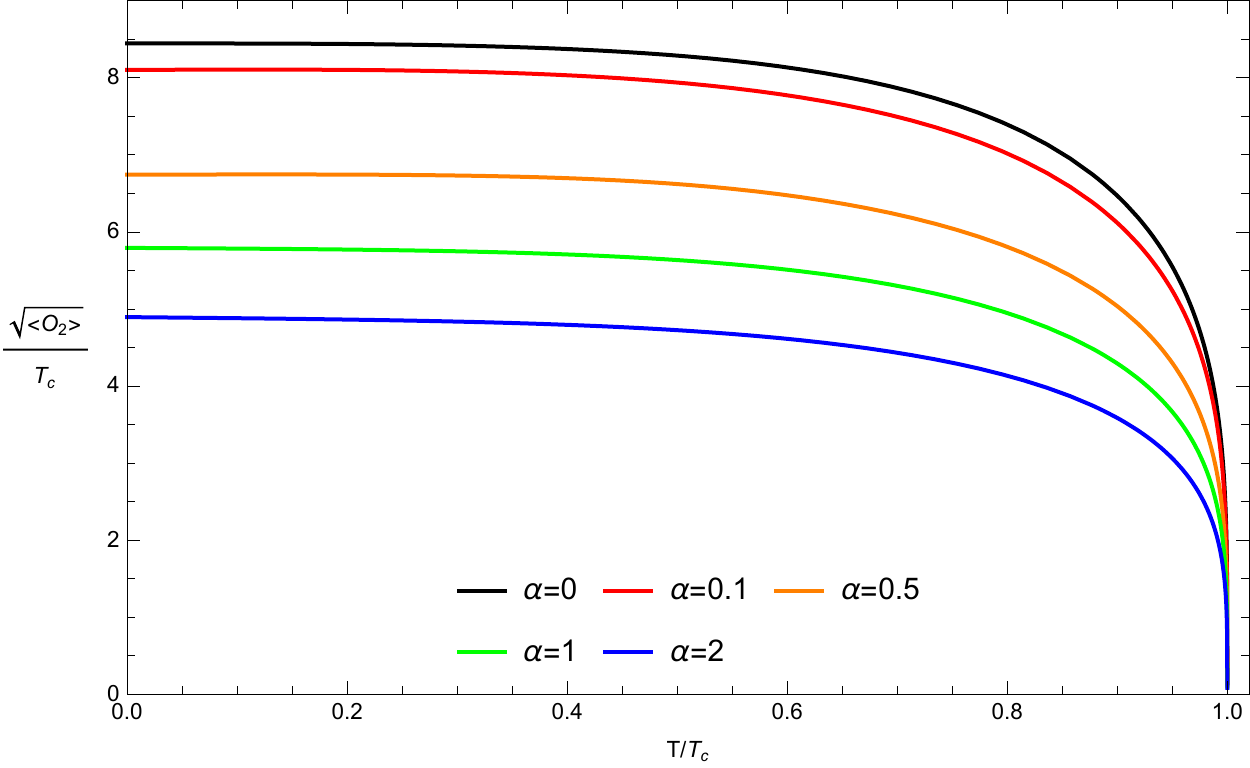}\caption{The
condensates of the scalar field as a function of the temperature $T/T_{c}$ for
various couplings $\alpha=0,0.1,0.5,1$ and $2$. \textbf{Left Panel:} The
condensate of the operator $\mathcal{O}_{1}$. \textbf{Right panel:} The
condensate of the operator $\mathcal{O}_{2}$.}%
\label{Fig2}%
\end{figure}

\begin{table}
	\begin{tabular}{|c||c|c|c|c|c|c|c|}
		\hline 
		$\alpha$ & 0 & 0.1 & 0.5 & 1 & 2 & 5 & 10\tabularnewline
		\hline 
		$2\times gap/T_{c}$ & 8.44 & 8.10 & 6.74 & 5.78 & 4.88 & 3.88 & 3.25\tabularnewline
		\hline 
	\end{tabular}
	
	\caption{The gaps of condensates $\left\langle \mathcal{O}_{2}\right\rangle $,
		which become smaller as $\alpha$ becomes larger.}
	
	\label{Table1}
\end{table}

In Fig. \ref{Fig2}, we plot the condensates $\left\langle \mathcal{O}%
_{1}\right\rangle $ and $\left\langle \mathcal{O}_{2}\right\rangle $ as a
function of the temperature $T$ for various couplings $\alpha$ in the left and
right panels, respectively. It is obvious that the condensates only occur
below the critical temperature $T_{c}$, and the curves with $\alpha=0$ recover
the results in \cite{Hartnoll:2008vx}. The condensates of different couplings
$\alpha$ have similar increasing trends with the decrease of temperature in
the both $\left\langle \mathcal{O}_{1}\right\rangle $ and $\left\langle
\mathcal{O}_{2}\right\rangle $ cases. Moreover, the condensate is smaller for
a larger $\alpha$, which implies that a stronger nonlinear coupling between
the scalar and electromagnetic fields makes the scalar hair easier to be
developed. Fitting the curves near the critical temperature $T_{c}$ for
$\left\langle \mathcal{O}_{1}\right\rangle $ and $\left\langle \mathcal{O}%
_{2}\right\rangle $, we find that the condensates behave as $\left\langle
\mathcal{O}_{i}\right\rangle \sim\left(  1-T/T_{c}\right)  ^{1/2}$ for all
values of $\alpha$. It is noteworthy that the condensate $\left\langle
\mathcal{O}_{1}\right\rangle $ diverges as $T\rightarrow0$, which means strong
backreactions on the background metric. Therefore, at extremely low
temperature, the probe limit is not a good approximation, and one may need to
solve the full Einstein equation with backreactions \cite{Hartnoll:2008kx}.
Unlike $\left\langle \mathcal{O}_{1}\right\rangle $, the condensate
$\left\langle \mathcal{O}_{2}\right\rangle $ approaches a constant as
$T\rightarrow0$. Actually, one can interpret the value of $\left\langle
\mathcal{O}_{2}\right\rangle $ at $T=0$ as twice the superconducting gap,
which is predicted to be $2\times$gap $=3.5T_{c}$ in BCS theory
\cite{Bardeen1957a}. We list the superconducting gaps for various couplings
$\alpha$ in Table 1, which shows that the superconducting gap becomes smaller
as the coupling $\alpha$ becomes larger, and the range of the gaps recovers
that of the high $T_{c}$ superconductors.

\subsection{Conductivity}

\begin{figure}[ptb]
\includegraphics[width=18cm]{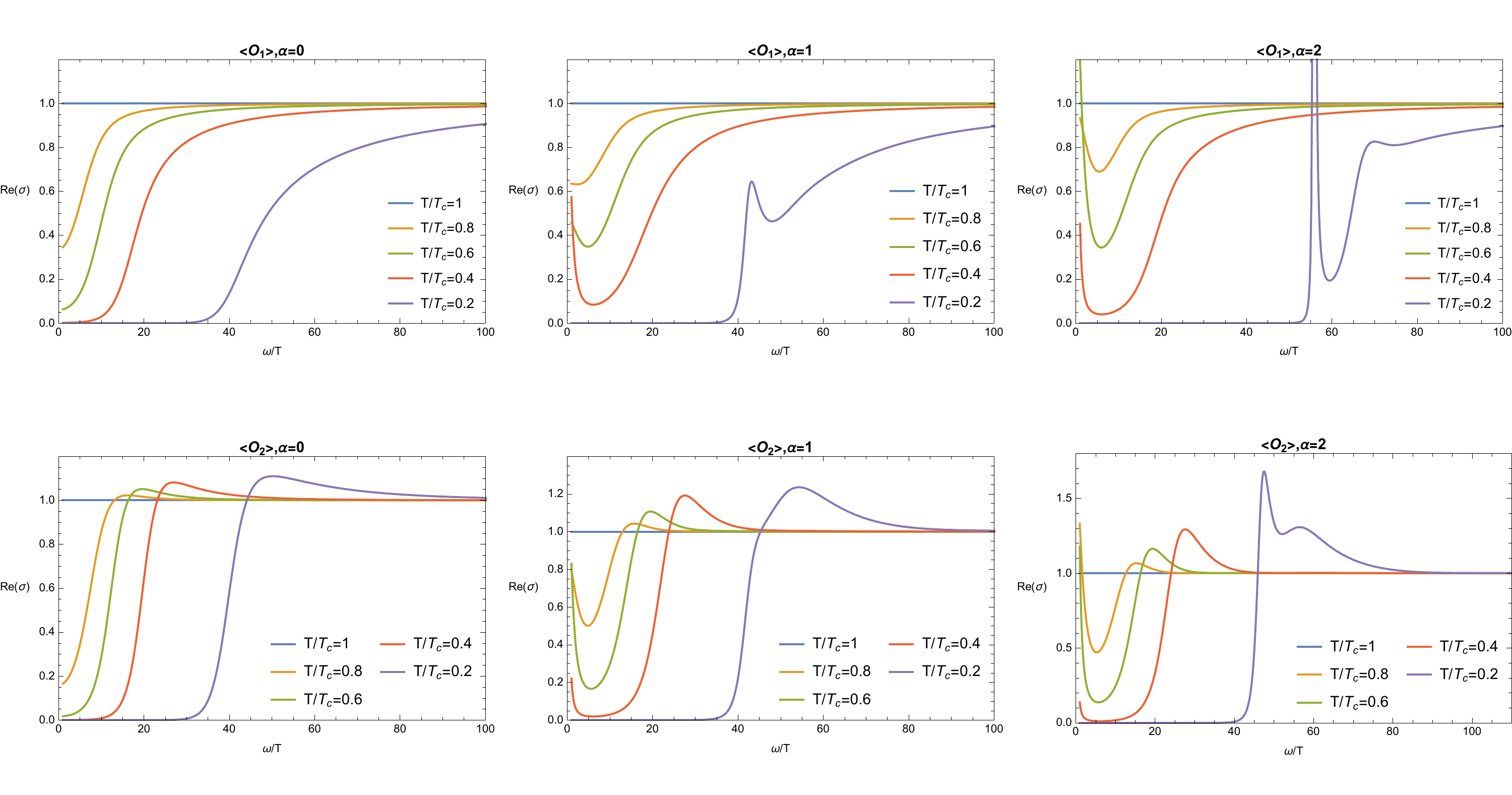}\caption{The real part of the
conductivity $\operatorname{Re}\left(  \sigma\right)  $ as a function of
$\omega/T$ for various $T/T_{c}$ with $\alpha=0$, $1$ and $2$. \textbf{Upper
row:} Condensating $\mathcal{O}_{1}$. \textbf{Bottom row:} Condensating
$\mathcal{O}_{2}$.}%
\label{Fig3}%
\end{figure}

\begin{figure}[ptb]
\includegraphics[width=18cm]{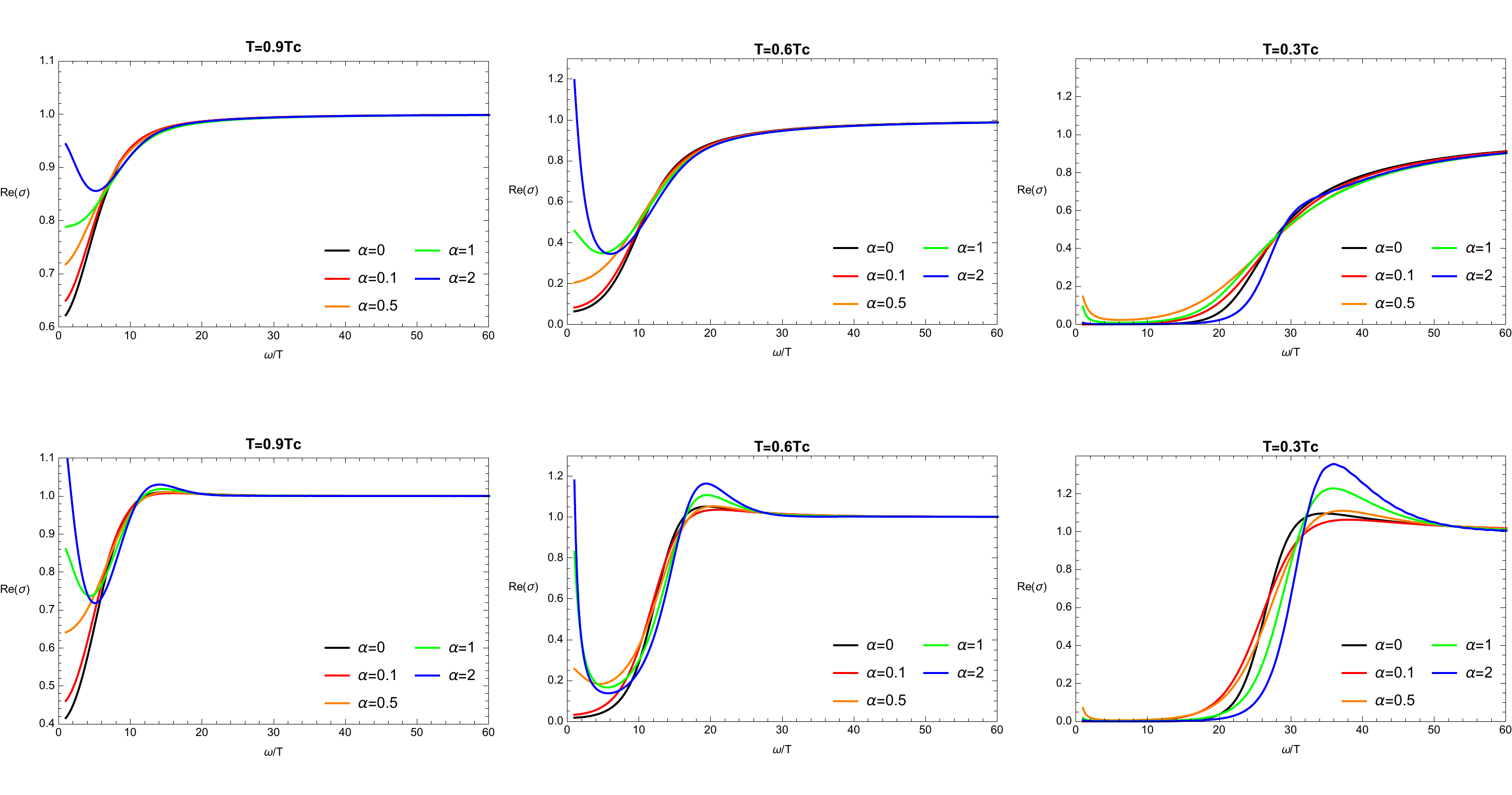}\caption{The real part of the
conductivity $\operatorname{Re}\left(  \sigma\right)  $ as a function of
$\omega/T$ for $\alpha=0,0.1,0.5,1$ and $2$ with a fixed temperature. The
upper and bottom rows depict $\operatorname{Re}\left(  \sigma\right)  $ for
condensating $\mathcal{O}_{1}$ and $\mathcal{O}_{2}$, respectively. The
temperatures from the left column to the right one are $0.9T_{c}$, $0.6T_{c}$
and $0.3T_{c}$, respectively.}%
\label{Fig4}%
\end{figure}

In this subsection, we investigate the optical conductivity in the dual CFT by
solving equation $\left(  \ref{eq:fluctuations' eq}\right)  $. In Fig.
\ref{Fig3}, we plot the real part of the conductivity $\operatorname{Re}%
\left(  \sigma\right)  $ versus the frequency of fluctuations with $\alpha=0$,
$1$ and $2$ in the both $\left\langle \mathcal{O}_{1}\right\rangle $ and
$\left\langle \mathcal{O}_{2}\right\rangle $ cases. Note that the left column
shows the $\alpha=0$ profiles, which recovers the results of
\cite{Hartnoll:2008vx}. The upper row of Fig. \ref{Fig3} displays the
conductivity for condensating $\mathcal{O}_{1}$, while the bottom one presents
the conductivity for condensating $\mathcal{O}_{2}$. The horizontal blue lines
correspond to the frequency-independent conductivity at the critical
temperature, which means that there is no condensate, and the system is dual
to a metal-like matter in the boundary. As one lowers the temperature below
$T_{c}$, a gap opens up and gets deeper. It is noteworthy that a spike appears
inside the gap at a low enough temperature. Particularly in the $\alpha=2$
case of condensating $\mathcal{O}_{1}$, the spike behaves like a delta
function, which will be discussed later. To better illustrate the effect of
the coupling $\alpha$ on the conductivity, we plot $\operatorname{Re}\left(
\sigma\right)  $ as a function of $\omega/T$ for a given $T/T_{c}$ in Fig.
\ref{Fig4}. In the upper/bottom row of Fig. \ref{Fig4}, the real part of the
conductivity is plotted for condensating $\mathcal{O}_{1}$/$\mathcal{O}_{2}$.
When the temperature is close to the critical temperature, e.g., $T=0.9T_{c}$,
the gap is shallower for a larger $\alpha$, indicating a larger
$\operatorname{Re}\left(  \sigma\right)  $. However, at a lower enough
temperature, the gap becomes significantly deep regardless of the values of
$\alpha$.

\begin{figure}[ptb]
\includegraphics[width=18cm]{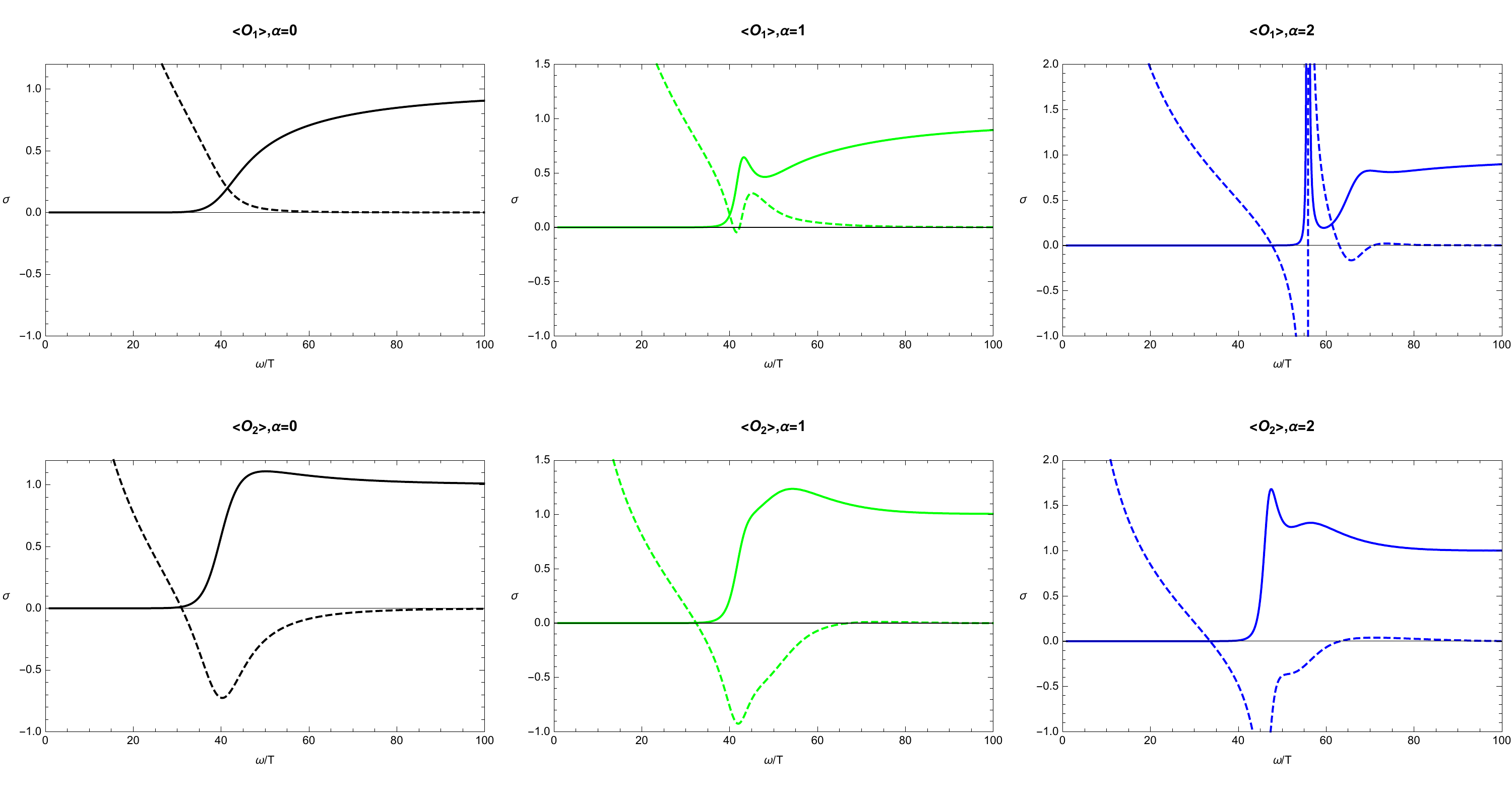}\caption{The real and imaginary parts
of the conductivity for $\alpha=0,1,2$ at the temperature $T=0.2T_{c}$. The
solid and the dashed lines represent the real and imaginary parts of the
conductivity $\sigma(\omega)$, respectively.}%
\label{Fig5}%
\end{figure}

It is well-known that $\operatorname{Re}\left(  \sigma\right)  $ has a delta
function at $\omega=0$, which is not plotted in the above figures. Although
the delta function can not be obtained directly from the numerical solution of
$\operatorname{Re}\left(  \sigma\right)  $, it can be verified from
$\operatorname{Im}\left(  \sigma\right)  $ according to the Kramers-Kronig relation%

\begin{equation}
\operatorname{Im}\left[  \sigma\left(  \omega\right)  \right]  =-\frac{1}{\pi
}\mathcal{P}\int_{-\infty}^{\infty}\frac{\operatorname{Re}\left[
\sigma\left(  \omega^{\prime}\right)  \right]  }{\omega^{\prime}-\omega
}d\omega^{\prime}.
\end{equation}
The real part of the conductivity has the form $\operatorname{Re}\left[
\sigma\left(  \omega\right)  \right]  \sim\pi n_{s}\delta\left(
\omega\right)  $, if the imaginary part of the conductivity has a pole,
$\operatorname{Im}\left[  \sigma\left(  \omega\right)  \right]  \sim
n_{s}/\omega$, where the coefficient $n_{s}$ is the superfluid density.
Therefore, one can roughly observe the poles by examining the numerical
results of the imaginary part of the conductivity as shown in Fig. \ref{Fig5}.
In Fig. \ref{Fig5}, we plot the real and imaginary parts of the conductivity
at a low temperature $T=0.2T_{c}$ by solid and dashed lines, respectively. The
real part of the conductivity all shows a deep gap, which can be characterized
by a gap frequency $\omega_{g}$, defined as the minimum point of the imaginary
part of the conductivity. In the unit of the critical temperature $T_{c}$, the
range of $\omega_{g}/T_{c}$ is approximately $8$ to $11$ for the three cases
in Fig. \ref{Fig5}. We also find that the gap frequency $\omega_{g}/T_{c}$ is
getting larger as the coupling $\alpha$ increases. Note that the spike occurs
in the case with the $\left\langle \mathcal{O}_{1}\right\rangle $ condensation
and $\alpha=2$. This spike behaves like a delta function and is dictated by a
pole of the imaginary part of conductivity. The occurrence of the spike
corresponds to the interference of reflected and incident waves when the
potential is high enough \cite{Horowitz:2008bn,Horowitz:2010gk}. Similar
spikes are also observed in other models \cite{Pan:2009xa,Pan:2010at}.

\section{Conclusions}

\label{sec:Conclusion}

In this paper, we investigated the holographic superconductor which is dual to
the EMS model coupled with a charged scalar field in the asymptotically AdS
spacetime. We focused on a non-minimal coupling function $h\left(
\Psi\right)  =e^{\alpha\Psi^{2}}$, which can lead to the spontaneous
scalarization of black holes. The properties of scalarized black holes have
been studied in \cite{Guo:2021zed,Guo:2021ere}. For simplicity, we studied the
holographic superconductor in the probe limit, which means the matter fields
do not backreact the background metric, but remains most of the interesting physics.

We first numerically solved the condensates of the scalar fields for the
operators $\mathcal{O}_{1}$ and $\mathcal{O}_{2}$, which are due to the choice
of the fall off $\Psi^{(1)}=0$ and $\Psi^{(2)}=0$, respectively. It showed
that the operators only condense when the temperature is below the critical
value $T_{c}$. By computing the critical temperatures $T_{c}$ of different
couplings $\alpha$, we found that the critical temperature $T_{c}$ grows with
the increase of $\alpha$. This result indicates that the effect of non-minimal
coupling can raise the critical temperature of the holographic superconductor,
which may provide an inspiration for high temperature superconductivity. We
next discussed the optical conductivity of the superconductor. The real part
of the conductivity is a constant when the temperature is above the critical
value $T_{c}$, which behaves like a metal in the normal phase. As the
temperature is lowered below the critical temperature, a gap is developed. One
can characterize the gap by the gap frequency $\omega_{g}/T_{c}$, which gets
larger as the coupling $\alpha$ increase. Interestingly, some spikes were
shown to occur in the gap for some large coupling $\alpha$ at a low
temperature. These spikes are associated with the interference of the
reflected wave and the incident wave when the potential is high enough.
Moreover, we found that the non-minimal coupling tends to make the spike occur
since the spike is observed at a higher temperature at a larger coupling
$\alpha$. Although the qualitative properties of holographic superconductors
can be obtained in the probe limit, it is always desirable to investigate
backreactions on the background metric. We leave this for future work.

\begin{acknowledgments}
We are grateful to Qingyu Gan and Guangzhou Guo for useful discussions and
valuable comments. This work is supported in part by NSFC (Grant No. 11875196,
11375121, 11947225 and 11005016), Special Talent Projects of Chizhou
University (Grant no. 2019YJRC001) and Anhui Province Natural Science
Foundation (Grant no. 1808085MA21).
\end{acknowledgments}

\bibliographystyle{unsrturl}
\bibliography{ref}

\end{document}